\documentclass[a4paper,final]{aipproc}

\layoutstyle{8x11single}

\usepackage{amsmath}
\usepackage{amssymb}
\usepackage{graphicx}

\newcommand{\OM}{\Omega_M}
\newcommand{\Om}{\Omega_m}
\newcommand{\Omo}{\Omega_m^0}

\newcommand{\OL}{\Omega_{\Lambda}}

\newcommand{\OLo}{\Omega_{\Lambda}^0}

\newcommand{\rc}{\rho_c}
\newcommand{\rco}{\rho_{c}^0}
\newcommand{\rmo}{\rho_{m}^0}

\newcommand{\rmr}{\rho_m}
\newcommand{\pmr}{p_m}

\newcommand{\wm}{\omega_m}

\newcommand{\rL}{\rho_{\CC}}

\newcommand{\rLo}{\rho_{\CC}^0}

\newcommand{\wD}{\omega_D}

\newcommand{\CC}{\Lambda}

\newcommand{\tOm}{\tilde{\Omega}_m}

\newcommand{\tOD}{\tilde{\Omega}_{D}}

\newcommand{\be}{\begin{equation}}
\newcommand{\ee}{\end{equation}}

\begin{document}

\title{Vacuum energy and cosmological evolution}


\author{Joan Sol\`a\footnote{sola@ecm.ub.edu}}{
  address={HEP Group, Departament d'Estructura i Constituents de la Mat\`eria and
Institut de Ci\`encies del Cosmos, Universitat de Barcelona, Diagonal
647, 08028 Barcelona, Catalonia, Spain.}}

\begin{abstract}
An expanding universe is not expected to have a static vacuum energy density. The so-called cosmological constant $\Lambda$ should be an approximation, certainly a good one for a fraction of a Hubble time, but it is most likely a temporary description of a true dynamical vacuum energy variable that is evolving from the inflationary epoch to the present day. We can compare the evolving vacuum energy with a Casimir device where the parallel plates slowly move apart (``expand''). The total vacuum energy density cannot be measured, only the effect associated to the presence of the plates, and then also their increasing separation with time. In the universe there is a nonvanishing spacetime curvature $R$ as compared to Minkowskian spacetime that is changing with the expansion. The vacuum energy density must change accordingly, and we naturally expect $\delta\Lambda\sim R\sim H^2$. A class of dynamical vacuum models that trace such rate of change can be constructed. They are compatible with the current cosmological data, and conveniently extended can account for the complete cosmic evolution from the inflationary epoch till the present days. These models are very close to the $\CC$CDM model for the late universe, but very different from it at the early times. Traces of the inherent vacuum dynamics could be detectable in our recent past.
\end{abstract}

\maketitle


\section{Introduction}

The cosmological term in Einstein's equations has traditionally been associated to a mysterious concept called the vacuum energy density: $\rL = \CC/(8\pi G)$. With the advent of quantum theory and quantum field theory (QFT) this association has been even more troublesome and has led to the famous cosmological constant (CC) problem\,\cite{CCproblem}. The CC problem is the main source of headache for every theoretical cosmologist confronting his/her predictions with the observational value of $\rL$\,\cite{CosmoObservations,CosmoObservations2,PLANCK1}. This is specially so after the discovery of a particle consistent with the Higgs boson at the LHC\,\cite{HiggsDiscovery}, as this may represent a serious endorsement to the notion of spontaneous symmetry breaking (SSB) and of electroweak vacuum energy in the context of the standard model (SM) of particle physics\,\cite{SM:EW,SM:QCD}. It is therefore more important than ever to properly address the notion of vacuum energy and its possible implications in cosmology. We will  elaborate on these important issues after considering the notion of dynamical vacuum energy and its interesting implications for a successful description of the entire cosmic history.

Despite its name, the possibility that the cosmological ``constant'' in Einstein equations is actually a time dependent quantity in cosmology, $\CC=\CC(t)$, is a most natural one, as it is difficult to conceive an expanding universe with a strictly constant vacuum energy density, $\rL$. It would  be tantamount to accept that such a quantity has remained immutable since the origin of time without having any theoretical explanation for this fact. A smoothly evolving vacuum energy density $\rL(t)=\rL(H(t))$ that inherits its time-dependence from the Hubble rate $H=H(t)$ as the natural dynamical variable in the Friedmann-Lema\^{\i}tre-Robertson-Walker (FLRW) background, is not only a qualitatively more plausible and intuitive idea, but is also suggested by fundamental physics, in particular by quantum field theory (QFT) in curved space-time\,\cite{BooksQFT} --  see also \cite{JSP-CCReview13} for a recent review on CC aspects of it.

In order to implement the notion of time evolving vacuum energy density is not necessary to resort to \textit{ad hoc} scalar fields, as usually
done in the literature (e.g. in quintessence formulations and the like). A ``running'' term can be expected on very similar grounds as one expects (and observes) the running of couplings and masses with a physical energy scale in QFT. Furthermore, the experimental evidence that the equation of state (EOS) of the dark energy (DE) could be evolving with time/redshift
(including the possibility that it might currently behave phantom-like) suggests that a time-variable term (possibly accompanied by a variable Newton's gravitational coupling
too, $G = G(t))$ could account in a natural way for all these features.

\section{Expansion dynamics and evolving vacuum energy}\label{sect:dynamicalvac}

The expanding universe can be treated as a perfect fluid. Let $\rho_{m}$ be the matter-radiation density  and $p_{m}$ the corresponding pressure. We denote by
$p_{m}=\omega_{m} \rho_{m}$ the equation of state (EoS) of that fluid. Recall that $\omega_{m}=0$ for
non-relativistic matter and $\omega_{m}=1/3$ for the radiation
component or relativistic matter. We can neglect the pressure deep in
the matter dominated era when the fluid behaves as dust.
The energy momentum tensor of matter reads ${T}_{\mu\nu}=
-p_{m}\,g_{\mu\nu}+(\rho_{m}+p_{m})U_{\mu}U_{\nu}$.
Let us write down Einstein's equations in the presence of the cosmological term:
\begin{equation}
R_{\mu \nu }-\frac{1}{2}g_{\mu \nu }R-\CC\,g_{\mu\nu}=8\pi G\ {T}_{\mu\nu}\,.
\label{EEsplit}
\end{equation}
The term $\CC$ is usually associated to the vacuum energy density
$\rho_{\Lambda}=\Lambda/(8\pi G)$.
The corresponding pressure of the vacuum ``fluid'' is $p_{\Lambda}=-\rho_{\Lambda}$.

The cosmological term $\Lambda\,g_{\mu\nu}$ on the
left hand side of Einstein's equations can be absorbed on the
right hand side with a modified energy-momentum tensor:
$\tilde{T}_{\mu\nu}\equiv T_{\mu\nu}+g_{\mu\nu}\,\rL $.
The above field equations can be given  the same form as in the case without CC term:
\begin{equation}
G_{\mu\nu}\equiv R_{\mu \nu }-\frac{1}{2}g_{\mu \nu }R=8\pi G\ \tilde{T}_{\mu\nu}\,,
\label{EE}
\end{equation}
with
$\tilde{T}_{\mu\nu}=-p_{\rm tot}\,g_{\mu\nu}+(\rho_{\rm tot}+p_{\rm
tot})U_{\mu}U_{\nu}$  the full energy momentum tensor including the effect of the
vacuum energy density. Here $\rho_{\rm tot}=\rho_{m}+\rho_{\Lambda}$ and
$p_{\rm tot}=p_{m}+p_{\Lambda}=p_{m}-\rho_{\Lambda}$. Explicitly:
\begin{equation}
\tilde{T}_{\mu\nu}= (\rho_{\Lambda}-p_{m})\,g_{\mu\nu}+(\rho_{m}+p_{m})U_{\mu}U_{\nu}\,.
\label{Tmunuideal}
\end{equation}
Hereafter we assume the FLRW metric with flat space slices. The two independent gravitational field equations read: \be
8\pi G (\rmr+\rL) = 3H^2 \;, \label{friedr} \ee
\be
 8\pi G (\wm\rmr-\rL) =-2{\dot H}-3H^2\;, \label{friedr2} \ee where
the overdot denotes derivative with respect to cosmic time $t$.

Despite a rigid CC term is the simplest possibility, let us emphasize
that the Cosmological Principle embodied in the FLRW metric still permits
the vacuum energy to be a function of the cosmic time, $\rL=\rL(t)$ or of other dynamical variables $\chi_i=\chi_i(t)$, i.e. $\rL=\rL(\chi_i(t))$. The same holds for the
gravitational coupling: $G=G (\chi_i(t))$.

In a cosmological context with dynamical parameters it is useful to consider the possible modifications that may undergo the basic conservation laws. The EoS for the vacuum energy density can still retain the usual form
$p_{\Lambda}(t)=-\rho_{\Lambda}(t)$.  The Bianchi identity satisfied by the Einstein tensor on the \textit{l.h.s.} of Eq.\,(\ref{EE}), namely $\nabla^{\mu}G_{\mu\nu}=0$, implies that the covariant derivative of the \textit{r.h.s.}  must be zero as well:
$\bigtriangledown^{\mu}\,\left(G\,\tilde{T}_{\mu\nu}\right)=0$, where the gravitational coupling $G$
is allowed to be variable. Using the explicit form of the
FLRW metric, the generalized conservation law emerging from this dynamical framework reads
\begin{equation}\label{BianchiGeneral}
\frac{d}{dt}\,\left[G(\rmr+\rL)\right]+3\,G\,H\,(1+\wm)\rmr=0\,.
\end{equation}
The concordance model, or  $\CC$CDM model\,\cite{Peebles1984}, appears as
a particular case of that relation in which the two parameters $\rL$ and $G$ stay constant. In such case, we recover the standard conservation law  $\dot{\rho}_m+3\,H\,(1+\wm)\rmr=0$, whose solution in
terms of the scale factor is well-known:
\begin{equation}\label{solstandardconserv}
\rmr(a)=\rmo\,a^{-3(1+\wm)}\,.
\end{equation}
The more general identity (\ref{BianchiGeneral}) suggest
possible extensions of the local conservation laws. They lead to the
following three classes of dynamical models:

Class I or $\CC_t$CDM model: $\CC=\CC(t)$ is assumed variable, and $G=$const. In this
case, Eq.\,(\ref{BianchiGeneral}) implies
\begin{equation}
\dot{\rho}_{m}+3(1+\omega_{m})H\rho_{m}=-\dot{\rho}_{\Lambda}\,. \label{frie33}
\end{equation}
Since in this case
$\dot{\rho}_{\Lambda}\neq 0$ it means we permit some energy exchange
between matter and vacuum, e.g. through vacuum decay into matter, or vice
versa.

Class II or $\CC_tG_t$CDM model: $\CC=\CC(t)$ is again variable, but $G=G(t)$ is also variable. In contrast to the previous case, here we assume matter conservation in the standard form (\ref{solstandardconserv}). As a result the following conservation law is projected:
\begin{equation}\label{Bianchi1}
(\rmr+\rL)\dot{G}+G\dot{\rho}_{\CC}=0\,.
\end{equation}
Here the evolution of the vacuum energy density is possible at the
expense not of an anomalous matter conservation law, but of a running gravitational coupling: $\dot{G}\neq 0$.

Class III or  $\CC G_t$CDM model: In this case we keep $\CC=$const., but $G=G(t)$ is again variable. Now we find:
\begin{equation}\label{dGneqo}
\dot{G}(\rmr+\rL)+G[\dot{\rho}_m+3H(1+\wm)\rmr]=0\,.
\end{equation}
This setup is different from the two previous ones, but shares some features: matter is non-conserved and the gravitational coupling is
running. Although the vacuum energy here is strictly constant, such situation can mimic a form of dynamical dark energy since it implies a different expansion rate.

Let us note that the above three generalized cosmological models can stay sufficiently close to the standard $\CC$CDM model.
It is therefore natural to ask what is the ``effective EoS'' of the dark energy for them\,\cite{SS05,BasSol2013}. By ``effective'' we mean here to reproduce the behavior of the above models from a setup where the
cosmological parameters $\CC$ and $G$ are assumed strictly constant, in
particular $\CC=0$, and where the DE is attributed to some smoothly
evolving, self-conserved, dynamical entity.  We will account for these effective EoS and study their implications.

\section{Time evolving vacuum versus running vacuum}\label{sect:timeversusrunning}
In the previous section we have described some useful frameworks for a possible time evolution of the the vacuum energy and the gravitational coupling. However, the generalized conservation law (\ref{BianchiGeneral}) in combination with Friedman's equation (\ref{friedr}) is not sufficient to determine the solution of the cosmological equations. We need additional input on the evolution laws $\rL=\rL(t)$ or $G=G(t)$. Although some suggestions have been given in the past on pure phenomenological grounds (cf. \,\cite{CCt Phenomenological, CCt Reviews} and references therein), we wish to focus here on a class of models more closely related
to QFT in curved spacetime\,\cite{ShapSol00,Fossil07,ShapSol09,Croats}\footnote{ See also the recent review \cite{JSP-CCReview13} and the long list of references therein.}.

We know that in  particle physics theories such as QED or QCD  the corresponding gauge coupling constants $g_i$ run with a scale $\mu$ that can be associated to the typical energy of the process, i.e. $g_i=g_i(\mu)$.
Similarly, in the effective action of QFT in curved spacetime $\rL$ can be thought of as an effective coupling depending on a mass scale $\mu$ associated to the cosmological evolution. In this context it is natural to assume that the running of $\rL$ from the quantum effects of the
matter fields is associated with the change of the spacetime curvature.
If so $\mu$ becomes tied to the typical energy of the classical gravitational external field of the FLRW metric. If we take into account that this energy is pumped into the quantum matter loops from the tails of the external gravitational
field, we can understand why it may lead to the physical running -- see \,\cite{JSP-CCReview13} for a more detailed discussion. Therefore we naturally associate $\mu^2$ to $R$, the scalar of curvature, which in the FLRW metric is a linear combination of $H^2$ and $\dot{H}$. For simplicity we adopt the simplest possible setting $\mu=H$, and we are led to the following generic form for the renormalization group (RG) equation for the vacuum energy density of the expanding universe\,\cite{ShapSol00,Fossil07,ShapSol09}:
\begin{eqnarray}\label{seriesRLH}
\frac{d\rL(H)}{d\ln
H^2}=\frac{1}{(4\pi)^2}\sum_{i}\left[\,a_{i}M_{i}^{2}\,H^{2}
+\,b_{i}\,H^{4}+c_{i}\frac{H^{6}}{M_{i}^{2}}\,+...\right]
\end{eqnarray}
Here $M_i$ are the masses of the fields involved in the loop contributions. It is clear that in Grand Unified Theories (GUT's) the heaviest fields available will provide the leading effect.  Let us point out that specific realizations of the structure (\ref{seriesRLH}) can be obtained in one-loop calculations within particular frameworks, see e.g. \,\cite{Fossil07}.

We should note that the terms $\sim M_i^4$ are excluded as  they would trigger a too fast running of $\rL$. In point of fact these terms are forbidden in our RG context since all known particles satisfy  $H<M_i$ and hence none of them is an active degree of freedom. In addition, only the even powers of $H$ are involved in the RG realization, owing to the general covariance of the effective action. The outcome is that the leading term for the present universe is the first one on the \textit{r.h.s.} of (\ref{seriesRLH}), i.e. the term proportional to $H^2$, whereas the next-to-leading term proportional to $H^4$ can only be relevant for the early universe where the value of $H$ can be very large. The terms of order $H^6$ or higher are suppressed by powers of the heavy masses, which are larger than $H$ in the RG decoupling regime.

If we integrate Eq.\,(\ref{seriesRLH}) to obtain
$\rL(H)$, an additive term (independent of $H$) obviously comes along. If we also retain the first two powers of $H$ in the result, we find that the dynamical cosmological term $\CC(H)=8\pi G \rL(H)$ takes on the following form:
\begin{equation}\label{lambdaH2H4}
\Lambda(H) = c_0 + 3\nu H^{2} + 3\alpha
\frac{H^{4}}{H_{I}^{2}} \;,
\end{equation}
where we have the two dimensionless coefficients
\begin{equation}\label{eq:nualphaloopcoeff}
\nu=\frac{1}{6\pi}\, \sum_{i=f,b} c_i\frac{M_i^2}{M_P^2}\,,\ \ \ \ \ \
\alpha=\frac{1}{12\pi}\, \frac{H_I^2}{M_P^2}\sum_{i=f,b} b_i\,,
\end{equation}
which play the role of $\beta$-function coefficients for the RG of the cosmological term. The sums in these formulas involve both fermions and bosons. There are also two dimensionful parameters, $c_0$ and $H_I$. The first one, $c_0$, is related (although not coincident) with the current value of the cosmological constant, $\CC_0$; and the second, $H_I$, is the energy scale associated to the inflationary expansion rate. This interpretation will become clear in the next section from the equations obtained from solving explicitly the model.

From the above RG formulation it is clear that the dimensionless
coefficients (\ref{eq:nualphaloopcoeff}) are predicted to be naturally small because $M_i^2\ll M_P^2$
for all the particles (even for the heavy fields of a typical GUT). For example, an estimate of $\nu$ within a generic GUT is found in the range
$|\nu|=10^{-6}-10^{-3}$\,\cite{Fossil07}. Similarly, the dimensionless coefficient $\alpha$ is expected small, $|\alpha|\ll 1$, because the inflationary scale $H_I$ should obviously be representative of the GUT scale $M_X$ below the Planck scale $M_P$. For example, if $M_X\sim 10^{16}$ GeV,  we have $M_X^2/M_P^2\sim 10^{-6}$.

Phenomenologically speaking, it is clear that in the context of the present universe $H^4\ll M_i^2\,H^2$, and the coefficient $\nu$ is the only one that can have an impact on the large scale structure and the Hubble expansion data. Its value can indeed be accessed by direct observations. Using a joint likelihood analysis of the recent supernovae type Ia data,
the CMB shift parameter, and the baryonic acoustic oscillations, one finds
that the best fit value for $\nu$ in the case of a flat universe is at
most of order $|\nu|={\cal O}(10^{-3})$\,\cite{BPS09}. This is also suggested by the analysis of cosmic perturbations\,\cite{FSS06}. In all cases the order of magnitude of $\nu$ is nicely in accordance with the aforementioned theoretical QFT expectations.

Equation (\ref{lambdaH2H4}) provides the necessary theoretical input to solve the dynamical vacuum models together with the remaining cosmological equations as outlined in the previous section. Let us mention that a generalization of the vacuum model  (\ref{lambdaH2H4}) in which the higher power $H^4$ is replaced by an arbitrary one $H^n\,(n>2)$ has also been studied\,\cite{LimaBasSola13a,HMention2013,Perico13} but the main results do not change with respect to the case $n=4$. In actual fact the model (\ref{lambdaH2H4}) is the natural one since it represents the minimal possible extension that is capable of supporting the complete cosmic history and remain compatible with the general covariance of the effective action of QFT in curved spacetime\,\cite{JSP-CCReview13}.

In the following we will use this type of vacuum models, and generalizations thereof, to generate explicit cosmological solutions with an eye at obtaining a description of the very early universe until the present one. In this way we can go beyond (and improve) the standard $\CC$CDM picture without disturbing its impeccable low energy success.

\section{Vacuum energy, Inflation and Graceful Exit}\label{sect:Graceful}

The Hubble function for the dynamical vacuum models under consideration can be derived by combining Eqs.(\ref{friedr}), (\ref{frie33}) and (\ref{lambdaH2H4}). We find the following differential equation:
\begin{equation}
\label{HE}
\dot H+\frac{3}{2}(1+\omega)H^2\left[1-\nu-\frac{c_0}{3H^2}-
\alpha\left(\frac{H}{H_I}\right)^{2}\right]=0.
\end{equation}

We note and remark the existence of two constant value solutions to this equation. On the one hand we have
$H=H_I[(1-\nu)/\alpha]^{1/2}$; it corresponds to the very early universe,
i.e. when $c_0\ll H^2$, and describes the de Sitter phase driven by a huge cosmological constant associated to the GUT scale $H_I\simeq M_X\gtrsim 10^{16}$ GeV. On the other hand, far away from the inflationary period ($H\ll H_I$) we
have $H=[c_0/3(1-\nu)]^{1/2}$, whereby $\Lambda\approx c_0\sim \Lambda_0$. This second constant value solution eventually leads to the late time cosmological constant behavior.

The solution of Eq.\,\eqref{HE} for $\omega=1/3$ and $c_0=0$ is
\begin{equation}\label{HS1}
 H(a)=\frac{\tilde H_I}{\left[1+D\,a^{4\,(1-\nu)}\right]^{1/2}}\,,
\end{equation}
 where we have defined $\tilde H_I \equiv \left(\frac{1-\nu}{\alpha}\right)^{1/2} H_I$. It is the critical
Hubble parameter associated to the initial de Sitter era.

For $D\,a^{4(1-\nu)}\ll1$ (during the very early universe) the solution
\eqref{HS1} can be approximated by the constant value  $H\approx
\tilde H_I$. In this period,  the vacuum energy density
remains essentially constant and coexists with a negligible radiation density, as we will check below. Such stage of almost constant vacuum energy density in the very early universe
obviously depicts the primeval de Sitter era in the cosmic evolution. In it the scale factor increases exponentially. This fact can be verified from (\ref{HS1}) by integrating from some initial (unspecified) scale factor up to a value $a$ well within the inflationary period, in which $D\,a^{4(1-\nu)}\ll1$ still holds good. The result is
\begin{equation}\label{eq:deSitter1}
a(t)\propto \exp\left\{\tilde H_I t\right\}\,,
\end{equation}
where the meaning of $t$ here is the elapsed time within the inflationary period.

After the inflationary phase is over, the term $D\,a^{4(1-\nu)}$ can be comparable to $1$ or much greater. Now the integration of Eq.\,(\ref{HS1}) must be done keeping that term, and we find
\begin{equation}\label{HS1b}
\int_{a_\star}^a\frac{d\tilde{a}}{\tilde{a}}\left[1+D\,\tilde{a}^{\,4\,(1-\nu)}\right]^{1/2}= \tilde H_I\,t\,,
\end{equation}
where $t$ here is  (in contrast to the previous case) the time elapsed after (approximately) the end of the
inflationary period, indicated by $t_\star$, and we have defined
$a_\star=a(t_\star)$. The integration constant $D$ is fixed from the
condition $H(a_\star)\equiv H_\star$, thus
\begin{equation}\label{eq:defD}
D=a_\star^{-4\,(1-\nu)}\left[\left(\frac{\tilde H_I}{H_\star}\right)^2-1\right]\,.
\end{equation}

At this point let us note that the ``graceful exit'' from the inflationary phase can be accommodated in this class of models\,\cite{LimaBasSola13a,HMention2013}.  Indeed, considering the limit $D\,a^{4(1-\nu)}\gg1$ of equation (\ref{HS1b}) we find
\begin{equation}
t\approx\,a^{2(1-\nu)}\,.
\end{equation}
As $|\nu|\ll 1$, it is obvious that after a sufficiently large value of the scale factor we have essentially
reached the radiation domination era, for which $a\propto
t^{1/2(1-\nu)}\simeq t^{1/2}$. We will reach the same conclusion below from the analysis of the energy densities in the transition period.

Using the Einstein equations (\ref{friedr}),(\ref{friedr2}) and the above solution for the Hubble function we can obtain the corresponding
vacuum and radiation energy densities in the early universe:
\begin{equation}\label{eq:rLa}
  \rho_\Lambda(a)=\tilde{\rho}_I\,\frac{1+\nu\,D\,a^{4(1-\nu)}}{\left[1+D\,a^{4(1-\nu)}\right]^{2}}\,,
\end{equation}
\begin{equation}\label{rho_1}
 \rho_r(a)=\tilde{\rho}_I\,\frac{(1-\nu)D\,a^{4(1-\nu)}}{\left[1+D\,a^{4(1-\nu)}\right]^{2}}\,.
\end{equation}
In the previous formulas we have defined
\begin{equation}\label{eq:rhoItild}
\tilde{\rho}_I\equiv\frac{3\tilde H_I^2}{8\pi G}\,,
\end{equation}
which is the primeval critical energy density associated with the initial de
Sitter stage. From (\ref{eq:rLa}) we see that $\tilde{\rho}_I$
provides the vacuum energy density for $a\to 0$.  We also note from the previous formulas that for $a\to 0$ we have
$\rho_r/\rL\propto a^{4(1-\nu)}\to 0$. It means, that the very early universe is indeed vacuum-dominated with a negligible amount of radiation.

\begin{figure}
\includegraphics[scale=0.57]{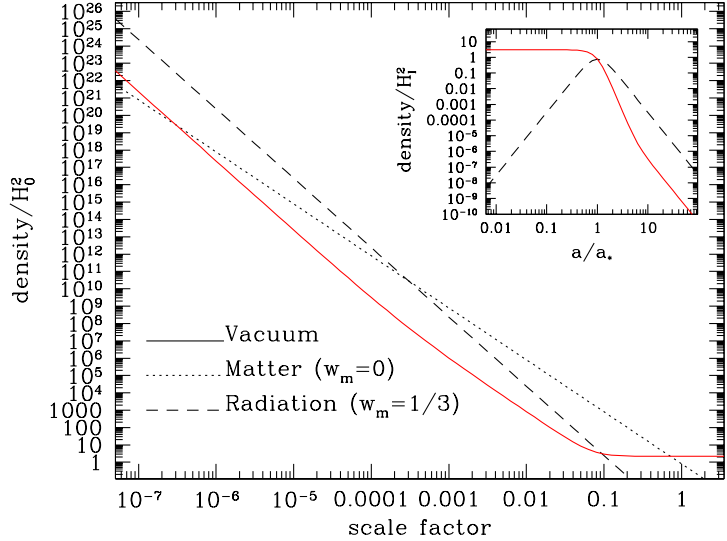}
  \caption{{\bf Outer plot}: The evolution of the radiation, matter and vacuum energy densities normalized with
respect to $H_0^2$, for the unified
vacuum model (\ref{lambdaH2H4})\,\cite{LimaBasSola13a,HMention2013,Perico13}. The curves shown are: radiation (dashed line), non-relativistic matter (dotted line) and vacuum (solid line, in red). Used inputs: $\nu=10^{-3}$, $\Omega^{0}_{m}=0.27$,
$\Omega^{0}_{R}=(1+0.227N_{\nu})\,\Omega^{0}_{\gamma}$,
$(N_{v},\Omega^{0}_{\gamma},h)\simeq (3.04,2.47\times 10^{-5}h^{-2},0.71)$\,\cite{CosmoObservations2}. {\bf Inner Plot:} The transition from the inflationary period into the FLRW radiation epoch. Same notation
for curves as before, although the densities are now normalized with
respect to $H_I^2$ and the scale factor with respect to the transition one $a_{\star}$ (see the text). We have set $\alpha=1$ for simplicity and used $8\pi G=1$ units.}
\label{Fig1}
\end{figure}

The aforementioned transition from the vacuum period into the radiation epoch (or ``graceful exit'' from the de Sitter phase)   is reconfirmed after inspecting the above energy densities. The solution for the radiation energy density \eqref{rho_1} acquires a
maximum value when the scale factor $a$ reaches the point
$a_{\star}\equiv D^{-1/4(1-\nu)}$, which is when the
inflation period is accomplished and the radiation-dominated era begins.

The analysis of the energy densities also shows that the vacuum energy becomes quickly subdominant and therefore does not interfere with the standard FLRW radiation regime. From
(\ref{eq:rLa}) and (\ref{rho_1}) the vacuum energy density is seen to follow a similar decay law as the radiation density,  $\rho_r\propto a^{-4(1-\nu)}\sim a^{-4}$, but is suppressed by the factor
$\rL/\rho_r\propto \nu$  (with $|\nu|\ll 1$). Such vacuum energy suppression at this epoch shows that the success of the BBN (Big Bang primordial Nucleosynthesis) is not compromised at all in this model!

The complete evolution of the matter and vacuum densities from the very early universe to the present day is displayed in Fig.\,\ref{Fig1}. In particular, the very early history is highlighted in the inner diagram of that figure, where we can clearly see the transition from the de Sitter vacuum-dominated phase into the radiation dominated phase.

The main conclusion from the above considerations is that the universe
with a vacuum energy density law (\ref{lambdaH2H4}) starts without a singularity and can overcome the horizon
problem. A light pulse beginning at $t=-\infty$ will have
traveled by the cosmic time $t$ a physical distance, $d_{H}(t)=
a(t)\int_{-\infty}^{t}\frac{d t'}{a(t')}$, which diverges thereby implying the absence of particle horizons. It follows that the local interactions may homogenize the whole universe.

Finally, let us note that the class of vacuum models (\ref{lambdaH2H4}) realize in an effective way the Starobinsky type of inflationary regime\,\cite{Starobinsky80} through the dominance of the highest power $H^4$  in the early universe, which is of order of the Starobinsky's correction term $R^2\sim H^4$. A generalized form of Starobinsky's inflation, based on the effective action of anomaly-induced inflation,  was considered long ago in\,\cite{Fossil07,ShapiroSola2002,JSola DESY2002,Russ02}. It was found that by taking the masses of the fields into account, the inflationary process automatically slows down and can therefore favor the graceful exit. At the end of the day, the class of vacuum models (\ref{lambdaH2H4}) encodes the basic features of different realizations of Starobinsky's type of inflation. This is specially welcome if we take into account the current good health of Starobinsky's inflation in the light of the most recent cosmological data and their implications on cosmological inflation\,\cite{PLANCK2,PLANCK3}.

\section{Generalized models for the running vacuum}\label{sect:Extended Models}

In the current and recent past, the role played by the  $H^4$ term in Eq.\,(\ref{lambdaH2H4}) is essentially irrelevant as compared to $H^2$ . Its sole ``raison d'\^etre'' was to provide the correct history of the very early universe and trigger a possible mechanism of inflation leading to graceful exit into the normal FLRW radiation phase. Once this phase is achieved we can continue the rest of the cosmological history with just the first two terms on the \textit{r.h.s.} of (\ref{lambdaH2H4}). Expressed as a vacuum energy density, we can write
\begin{equation}\label{eq:VacuumfunctionCC}
\rho_{\Lambda}=n_0 +n_2 H^2\,,
\end{equation}
where $n_0$ is a dimensionful constant that is related to the current vacuum energy density, and $n_2$ is another dimensionful constant proportional to $\nu$.
It is convenient to normalize the previous formula as follows:
\begin{equation}\label{QFTModelLowEnergy}
\rho_{\Lambda}(H)=\rLo+\frac{3\,\nu}{8\pi}\,M_P^2\,(H^2-H_0^2)\,,
\end{equation}
where  $M_P$ is the Planck mass (related to the gravitational coupling through $M_P=G^{-1/2}$) and $H_0$ is the current Hubble rate. The previous normalization implies that $\rho_{\Lambda}(H_0)=\rLo$ (the current vacuum energy density). The remaining $H^2$ dependence in (\ref{QFTModelLowEnergy}) carries some slow vacuum dynamics even today. We can fit this dynamics -- encoded in the dimensionless parameter $\nu$ -- by comparing the model with observations. From a joint likelihood analysis of the recent supernovae type Ia data, the CMB shift parameter, and the baryonic acoustic oscillations one finds that the
best fit parameters for a flat universe are: $\Omega_{m0}\simeq 0.27-0.28$
and $|\nu|={\cal O}(10^{-3})$ (see \cite{BPS09}). It is
encouraging to realize that the fitted value of $\nu$ is within the theoretical expectations when this parameter is identified to the $\beta$-function of the running CC within QFT in curved spacetime. As already mentioned, in specific QFT frameworks one typically finds $\nu=10^{-5}-10^{-3}$ \cite{Fossil07}. Let us also mention recent alternative QFT frameworks suggesting a
similar kind of evolution of the vacuum energy leading once more to
dynamical terms $\sim H^2$\,\cite{Maggiore,Bilic,Hack13}.

For the recent universe we set $\alpha=0$ in (\ref{HE}), and also $\wm=0$ (corresponding to the matter-dominated epoch). Since the vacuum energy density is variable the Bianchi identity can be satisfied in different ways, as explained in the first section after the introduction. Let us adopt the Class I or $\CC_t$CDM model defined in Eq.\,(\ref{frie33}). After some calculations the Hubble function and corresponding energy densities can be readily derived as a function of the scale factor. For example, the Hubble function reads:
\begin{equation}
\label{eq:Ha1}
{H}^{2}(a) = \frac{H_0^2}{1-\nu} \left[(1-\Omega_{\CC}^{0})\,a^{-3(1-\nu)}+\Omega_{\Lambda}^0-\nu \right]\,.
\end{equation}
Here $\Omega_{\Lambda}^0$ is the standard cosmological parameter associated to the cosmological term. The superindex ``$0$'' in it denotes the current value.  Similarly, we obtain the matter density:
\begin{equation}\label{mRGa1}
\rho_m(a) =\rho_m^0\,a^{-3(1-\nu)}\,,
\end{equation}
and the vacuum energy density:
\begin{equation}\label{CRGa1}
\rL(a)=\rLo+\frac{\nu\,\rho_m^0}{1-\nu}\,\left[a^{-3(1-\nu)}-1\right]\,.
\end{equation}
As expected, $\rho_m(a)$ and  $\rL(a)$ yield the current values $\rho_m^0$ and  $\rLo$ for $a=1$. Moreover, the above energy densities boil down to their respective $\CC$CDM forms for $\nu=0$, namely the matter density reduces then to Eq.\,(\ref{solstandardconserv}) (with $\omega_m=0$ in the matter dominated epoch), and the vacuum energy density stays constant at $\rLo$.

Let us also note from Eq.\,(\ref{eq:Ha1}) that $H(a)$ takes the standard form for $\nu=0$, that $H(a=1)=H_0$ for any $\nu$, and that in the very late universe we get an effective cosmological constant dominated era,
$H\approx H_0\,\sqrt{(\Omega_{\Lambda}^0-\nu)/(1-\nu)}$, that implies a a new pure de Sitter phase. This is the late-time de Sitter phase or
DE epoch.

The model (\ref{QFTModelLowEnergy}) can be generalized, still within the Class I of dynamical vacuum models, by including a term which is proportional to $\dot{H}\equiv dH/dt$. Such term was mentioned before, it has the same dimension as $H^2$ and is equally compatible with the general covariance of the effective action. The generalized vacuum model reads:
\begin{equation}\label{eq:VacuumfunctionCCtCDM}
\rho_{\Lambda}(H,\dot{H})=n_0 +n_2 H^2 +n_{\dot{H}}\,{\dot H}\,.
\end{equation}
We point out that for all values of $(n_2,n_{\dot{H}})$ it is essential that $n_0\neq 0$ for these models to work. If $n_0=0$, we find that the universe cannot have a transition point from deceleration to acceleration, for all models of the form (\ref{eq:VacuumfunctionCCtCDM}), and at the same time the predictions on structure formation are quite odd -- see \,\cite{BPS12,BasSol14} for details, and also \cite{Japanese12,Japanese3}. One also finds that with $n_0=0$ the model can partially solve the background problem if we replace $\dot{H}\to H$ in Eq.\,(\ref{eq:VacuumfunctionCCtCDM}), but the structure formation problems remain --- see \cite{BasSol14} for a most recent study. So, $n_0\neq 0$ is mandatory, and this fact excludes the entire class of the so-called ``entropic-force dark energy models''\,\cite{Easson10,Easson10b} -- not to be confused with the general approach to entropic gravity as formulated e.g. in\,\cite{EntropicG1,EntropicG2}.

The solution of these generalized vacuum models is also completely analytical. Let us reparameterize the expression (\ref{eq:VacuumfunctionCCtCDM}) conveniently. Define the coefficient $\nu$ as before, and introduce the new dimensionless coefficient $\eta$  as follows:
\begin{equation}\label{eq:defn2ndH}
n_2=\frac{3\nu}{8\pi}\,M_P^2\,, \ \ \ \ \ \ \ n_{\dot{H}}=\frac{\eta}{4\pi}\,M_P^2\,.
\end{equation}
It is obvious that for $\eta=0$ we recover the model (\ref{eq:VacuumfunctionCC})-(\ref{QFTModelLowEnergy}). The  Hubble function for the generalized model in the matter dominated epoch can be  derived, and provides the result:
\begin{equation}\label{eq:HzCCtCDM}
H^2(a) = H_0^2\left[\frac{\Omo}{\xi}~a^{-3 \xi}
                           + \frac{\OLo-\nu}{1 - \nu}\right]\,,
\end{equation}
where
\begin{equation}\label{eq:defxi}
\xi  = \frac{ 1 - \nu }{ 1 - \eta }\,.
\end{equation}
After a straightforward calculation the matter and vacuum energy densities can also be determined:
\begin{equation}\label{eq:MatterdensityCCtCDM}
\rho_m(z) =  \rmo~a^{-3 \xi}
\end{equation}
and
\begin{equation}\label{eq:CCdensityCCtCDM}
\rL(z)=\rLo+{\rmo}\,\,(\xi^{-1} - 1) \left[ a^{-3\xi} -1  \right]\,,
\end{equation}
For $\eta=0$ these expressions boil down to (\ref{mRGa1}) and (\ref{CRGa1}) respectively, as they should.

Next we discuss the Class II of models, i.e. the  $\CC_tG_t$CDM models. The basic differential equation in this case is Eq.\,(\ref{Bianchi1}). Here
matter is covariantly conserved and the Bianchi identity can be fulfilled
through a dynamical interplay between a running vacuum energy and a
running gravitational coupling. Unfortunately,
an explicit analytical solution of the cosmological equations is not possible in this case, if we adopt a vacuum dynamical law of the generalized form as indicated in
Eq.\,(\ref{eq:VacuumfunctionCCtCDM}), i.e. containing both $H^2$ and
$\dot{H}$ terms. Since, however, the terms $H^2$ and $\dot{H}$ play a similar role we will focus here on the model type
(\ref{eq:VacuumfunctionCCtCDM}) under the assumption that $n_{\dot{H}}=0$
(i.e $\eta=0$)\,\cite{GSFS10}. As we are left once more with the single dimensionless parameter $\nu$, a more amenable treatment is possible.

The running gravitational coupling can then be explicitly computed as a logarithmic function of the Hubble rate:
\begin{equation}\label{GH}
G(H)=\frac{G_0}{1+\nu\,\ln{H^2/H_0^2}}\,,
\end{equation}
where $G_0=1/M_P^2$ is the current value of Newton's gravitational constant. The logarithmic behavior indicates that $G$ varies very slowly with the cosmic evolution. If we focus on the nonrelativistic epoch, the corresponding solution in terms of the scale factor can be
obtained in the form of an implicit equation:
\begin{eqnarray}\label{eq:fMCCtGtCDM}
\frac{1}{g(a)}-1+\nu\,\ln\left[\frac{1}{g(a)}-\nu\right]
=\nu\,\ln\left[\Om(a)+\OLo-\nu\right]\,.
\end{eqnarray}
Here $g(a)$ is the gravitational coupling at an arbitrary value of the scale factor, normalized to its current value $G_0$,i.e. $g(a)=G(a)/G_0$. As expected, $g(a)=1$ for $\nu=0$. We have defined $\Om(a)=\rmr(a)/\rco=\Omo\,a^{-3}$, i.e. the matter energy density
normalized to the current critical density. The dynamical vacuum energy density in this model can be expressed as follows:
\begin{eqnarray}\label{eq:fLCCtGtCDM}
\rL(a)=\frac{\rLo}{1-\nu\,g(a)}\,\left\{1+\frac{\nu}{\OLo}\,\left[\Om(a)\,g(a)-1\right]\right\}\,,
\end{eqnarray}
where $g(a)$ is the implicit function defined in (\ref{eq:fMCCtGtCDM}). Notice that for $\nu=0$ we have $\rL=\rLo$, as it should be.

Finally, let us also consider the Class III of dynamical models, which obey the local conservation law (\ref{dGneqo}). Here the vacuum
term is not evolving with the expansion, but the effective description of
this scenario leads to a case of virtually dynamical DE. The effect is
caused by the fact that matter is not conserved in such model, and
this can be covariantly consistent provided there is a small running of
the gravitational coupling, resulting in a
a nontrivial evolution of the effective DE\,\cite{BasSol2013}. To illustrate the present
class, we can retake the anomalous evolution law for dust matter, Eq.(\ref{mRGa1}), which presents a small departure from the standard law (\ref{solstandardconserv}). We write the anomalous conservation law as follows:
\begin{equation}\label{nonstandardmatterdelta}
\rmr(a)=\rmo\,a^{-3(1-\varepsilon)}\,,
\end{equation}
where $|\varepsilon|\ll 1$ is a small parameter. We do not denote it $\nu$ this time to emphasize that there is no correlation in this case with a time variable vacuum energy.

Trading the cosmic time by the scale factor, the differential equation (\ref{dGneqo}) adopts the following form:
\begin{equation}\label{dGneqo2}
G'(a)\left[\rmr(a)+\rLo\right]+G(a)\left[{\rho}'_m(a)+\frac{3}{a}\,\rmr(a)\right]=0\,.
\end{equation}
The primes indicate differentiation with respect to the scale factor. From the anomalous matter conservation law (\ref{nonstandardmatterdelta}) we can solve Eq.\,(\ref{dGneqo2}) for $G(a)$. We find\,\cite{FritzschSola2012}:
\begin{equation}\label{GafixedL}
G(a)=G_0\,\left[\Omo\,a^{-3(1-\varepsilon)}+\OLo\right]^{\varepsilon/(1-\varepsilon)}\,.
\end{equation}
Here $G_0=1/M_P^2$ is the current value of Newton's constant.
The corresponding Hubble function can be written in terms of the evolving gravitational coupling (\ref{GafixedL}) as follows:
\begin{equation}\label{GrelatedH}
\frac{H^2(a)}{H_0^2}=\left[\Omo\,a^{-3(1-\varepsilon)}+\OLo\right]^{1/(1-\varepsilon)}=\left[\frac{G(a)}{G_0}\right]^{1/\varepsilon}\,.
\end{equation}
As a result, for small $|\varepsilon|\ll 1$, the gravitational coupling $G$ runs logarithmically with the Hubble rate:
\begin{equation}\label{GzfixedLsmallnu}
G(H)\simeq G_0\,\left(1+\varepsilon\,\ln\frac{H^2}{H_0^2}+{\cal
O}(\varepsilon^2)\right)\,.
\end{equation}
At leading order in  $\varepsilon$ this expression for the variation of $G$ is
very similar to Class II considered previously, see Eq.\,(\ref{GH}), except that here the vacuum energy is not running. In actual fact, $\varepsilon\simeq -\nu$ for small values of these parameters.

\section{Effective equation of state analysis: quintessence- and phantom-like behavior}\label{sect:EoS Analysis}

We start from the Hubble function for a general vacuum framework in flat space:
\begin{eqnarray}\label{eq:Hgeneral}
{H^2(z)}=\frac{8\pi\, G(z)}{3}\left[\rmr(z)+\rL(z)\right]\,.
\end{eqnarray}
Here we take into account the possible cosmic evolution of
both $G$ and $\rL$ as a function of the scale factor or equivalently of the redshift $z=(1-a)/a$. Any of the models considered in the previous sections can be put in the above form. It is convenient to introduce the current values of the cosmological parameters associated to the generic model (\ref{eq:Hgeneral}). We denote them $\Om^0=\rmr^0/\rc^0$ and $\OL^0=\rL^0/\rc^0$. Since we assumed flat space, we have $\Om^0+\OL^0=1$.

The ``effective'' EoS parameter $\wD$ for the generalized vacuum model
introduced above is obtained by equating the expansion rate (\ref{eq:Hgeneral}) to that of a self-conserved entity with
negative pressure, i.e.
\begin{equation}\label{H2DE}
H_{\rm D}^2(z)=H_0^2\,\left[\tOm^0\,(1+z)^{3}+\,\tOD^0\,\zeta(z)\right]\,,
\end{equation}
where
\begin{equation}\label{eq:defzz}
\zeta(z)\equiv\,\exp\left\{3\,\int_0^z\,dz'
\frac{1+\wD(z')}{1+z'}\right\}\,.
\end{equation}
The expansion rate (\ref{H2DE}) defines the ``DE picture'' of the original dynamical vacuum model (\ref{eq:Hgeneral})\,\cite{SS05}. In this picture, the cosmological parameters are $\tOm^0$ and $\tOD^{0}=1-\tOm^0$.

Let us emphasize that the cosmological parameters in the two pictures
need not be identical; in particular, the value of the differential mass parameter
\begin{equation}\label{eq:DeltaOmega}
\Delta\Om^0\equiv \Om^0-\tOm^0\,,
\end{equation}
even if it is naturally expected small ($|\Delta\Om^0|/\Om^0\ll 1$),  can
play a role in our analysis.

\begin{figure}
\includegraphics[scale=0.57]{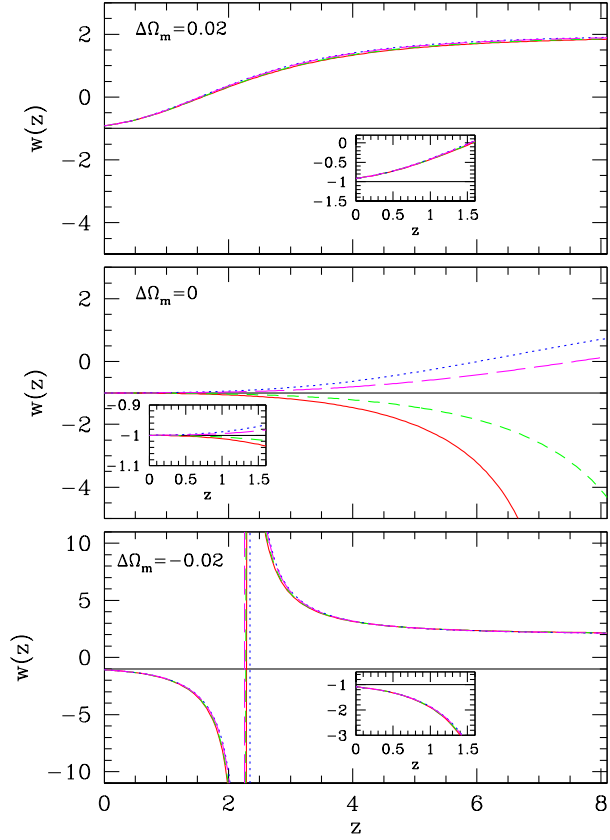}
\caption{
{\em Outer Plots:} The evolution of the effective EoS parameter
for various $\Delta \Omega_{m}^{0}$ values within the $\CC_t$CDM (Class I) model. We assume $\eta= 3\nu/4$.
The various curves correspond to
$\nu=-0.002$ (blue/dotted), $\nu=-0.001$ (magenta/long dashed),
$\nu=0.001$ (green/dashed) and $\nu=0.002$ (red/solid).
{\em Inner Plots:}
Detail of the evolution of $w_{D}(z)$ in the shorter range  $0\leqslant z\lesssim 1.5$ relevant for SNIa observations\,\cite{BasSol2013}.}
\end{figure}

The next point is to implement an important matching condition between the
two expansion histories, namely we require the equality of the expansion
rates of the original dynamical ``CC picture'' (\ref{eq:Hgeneral}) and that of
the DE picture (\ref{H2DE}): $H(z)=H_{\rm D}(z)$ ($\forall z$). First of
all we note from equations (\ref{H2DE}) and (\ref{eq:defzz}) that
$\wD(z)=-1+(1/3)\,[({1+z)}/{\zeta}]\,{d\zeta}/{dz}$. From here a
straightforward calculation leads us to a first operative formula to
compute the effective EoS:
\begin{eqnarray}\label{wDE1}
\wD(z)=-1+
\frac{\frac13\,(1+z)\,{dE^2(z)}/{dz}-(1+z)^{3}\,\tOm^0}{E^2(z)-\tOm^0\,(1+z)^{\alpha_m}}\,.
\end{eqnarray}
It is important to remark that the normalized Hubble rate  $E(z)\equiv H(z)/H_0$ in the above EoS formula must be computed
from (\ref{eq:Hgeneral}), in which the matter and vacuum density functions are assumed to be known from the structure of the given
dynamical vacuum model.

We can readily check that for the special case of the $\CC$CDM model,
and for $\Om^0=\tOm^0$,
Eq.\,(\ref{wDE1}) reduces to $\wD=-1$, as expected. It follows that any
departure from this result will be a clear sign that the background
cosmology cannot be one with $\CC=$const. In particular, if the resulting
effective EoS evolves with the expansion, $\wD=\wD(z)$, it will be a sign
of a dynamical vacuum.

Equation (\ref{wDE1}) can be explicitly worked out for the Class I, II and III models described in the previous sections. For example, if we take the Class I model (\ref{eq:VacuumfunctionCCtCDM}) and expand the result linearly in the small
parameters $\nu$ and $\eta$, assuming that $z$ is not very large (i.e.
for points relatively close to our current universe) and with the natural
assumption $|{\Delta\Omo}/{\Omo}|\ll1$, we arrive at the following approximate effective EoS:
\begin{equation}\label{eq:EoSCCtCMDparamsmall}
\wD(z)\simeq
-1+\frac{\Omo}{\OLo}(1+z)^3\left[\frac{\Delta\Omo}{\Omo}+3(\eta-\nu)\,\ln(1+z)\right]\,.
\end{equation}
This approximate equation comprises all the main qualitative features of the effective EoS. Similar equations can be computed for the other models -- see \cite{BasSol2013} for details -- but
let us focus on the particular case (\ref{eq:EoSCCtCMDparamsmall}). As expected we can recover the $\CC$CDM limit, $\wD(z)=-1$, for
$\nu=\eta=0$ and $\Delta\Omo=0$. At the same time for $\Delta\Omo=0$ we get $\wD=-1$ at $z=0$
irrespective of $\nu$ and $\eta$.

Consider now the ostensible quintessence- and phantom-like behaviors encoded in (\ref{eq:EoSCCtCMDparamsmall}). Once more we take $\Delta\Omo=0$ to display the desired behavior only in terms of the genuine parameters of the model, $\eta$ and $\nu$. The effective EoS indeed mimics quintessence
(viz. $\wD\gtrsim -1$) for $\eta>\nu$, and phantom DE (viz. $\wD\lesssim-1$) for $\eta<\nu$.

\begin{figure}
\includegraphics[scale=0.57]{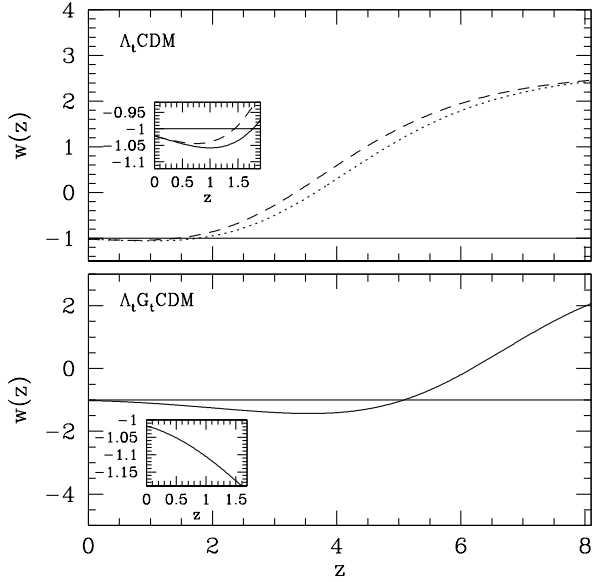}
\caption{
{\em Upper Panel:}  The effective EoS parameter for the $\CC_t$CDM model (Class I model). It is assumed a
future accuracy $\Delta\Omega_m^0=-0.005$ and two cases:
$\eta=-\nu=3\times 10^{-3}$ (dashed line) and $\eta=-3\nu/4$, with $\nu=-3\times 10^{-3}$ (dotted line).
{\em Lower Panel:} Effective EoS parameter versus $z$ for the alternative
$\CC_t$$G_t$CDM model (Class II) under the same assumptions as before. Here it is assumed $\eta=0$
and  $\nu=-3\times 10^{-3}$.
{\em Inner Plots:}
The corresponding EoS parameter for a redshift interval highly focused near out time\,\cite{BasSol2013}.}
\end{figure}

Obviously, if there is a significant deviation
between the parameters $\Omo$ and $\tOm^0$ the situation could change,
depending on the size and sign of the term ${\Delta\Omo}/{\Omo}$.
The numerical analysis of the exact EoS can be performed for given input parameters using Class I models (Fig. 2) and Class I and Class II models (Fig. 3). We verify that with the current values of the cosmological parameters and their precisions we get $\wD=-1\pm0.1$ for supernovae data
at small redshifts $z<1$ (cf. once more Fig. 2 and 3).

Let us also point out a curious effect displayed by the exact effective EoS for relatively large redshifts of order $\sim 2$ away from our recent past, and hence not at reach of the approximation used to derive Eq.\,(\ref{eq:EoSCCtCMDparamsmall}). It can be observed in Fig.\,2.  It is a kind of
divergent behavior due to the fact that the denominator of
Eq.(\ref{wDE1}) vanishes around $z\simeq 2$ for the Class I model under consideration. In point of fact there is no real singularity here because the energy densities (\ref{eq:MatterdensityCCtCDM}) and (\ref{eq:CCdensityCCtCDM}) of the
underlying vacuum model behave smoothly for all values of the
redshift. It is only the effective EoS
description of the model that leads to this fake
singularity, which is nothing but an artifact of the
reparametrization of the running vacuum model as if it were a DE
 model with expansion rate (\ref{H2DE})\,\footnote{Similar effects have bee reported in the literature, for example when one reconstructs the equation of state of dark energy,
 see e.g. \cite{Sahni2008}.}. It goes without saying that if we would identify such sort of anomaly in the observational data, we could
 suspect that there is no fundamental quintessence-like field behind the
 EoS but something else. It could be a signature of the dynamical vacuum model under consideration!

Let us finally mention that the effective EoS of the dynamical vacuum model does not adapt to the simple parameterizations of the DE in vogue in the literature\,\cite{CPL}. These simple parameterizations cannot capture the dynamic features of any of the Classes I, II and III of vacuum models we have considered, and in this sense they have a limited use -- see \,\cite{BasSol2013} for a more elaborated discussion. Our analysis, therefore, shows the need to remain open minded to more general EoS forms, as they could encode the truly fundamental behavior of the quantum fields in curved spacetime.

\section{Cosmic time evolution of Fundamental ``Constants''}\label{sect: Fundamental Constants}

Perhaps one of the most important motivations for the dynamical vacuum energy in the universe is that it may provide an explanation for the possible time evolution of the fundamental constants, in particular the particle masses. This has recently been emphasized in \cite{FritzschSola2012} and \cite{FritzschSola2014}.

The potential cosmic time evolution of the masses in particle physics is intriguingly connected with the frequent hints of the time variation of the fundamental constants -- see \cite{Fritzsch11,CalmetFritzsch06}. For comprehensive reviews, cf. \cite{FundConst}.
In our context we can find a possible explanation. Namely, in order to preserve the Bianchi identity that is satisfied by the Einstein tensor of the gravitational field equations (\ref{EE}) ($\nabla^{\mu}G_{\mu\nu}=0$), the time evolution of the masses can be compensated for by the time variation of one or more fundamental gravitational parameters, typically the gravitational constant $G$, or the cosmological constant $\CC$, or both. So, in this proposal, dynamical dark energy is naturally implied as a conceivable fundamental explanation for the observational hints on the variation of the fundamental constants.

We can briefly sketch a theoretical scenario to implement that idea. Take e.g. the baryonic density in the universe, which is essentially the mass density of
protons. We can write $\rho_p=n_p\,m_p$, where $n_p$ is the number density
of protons and $m_p^0=938.272 013(23)$ MeV is the current proton mass. If
the mass density is non-conserved, it may be due to the fact that the proton mass $m_p$ does not stay strictly constant with time and
scales mildly with the cosmic evolution:
\begin{equation}\label{eq:nonconservedmp}
m_p(a)=m_p^0\,a^{3\nu}\,, \ \ \ \ \ (|\nu|\ll 1)\,.
\end{equation}
Combining this equation with $n_p=n_p^0\,a^{-3}$ (the normal particle number dilution law associated to the cosmic expansion, with $n_p^0$ the current number density of protons) we find that the proton density at any time is $\rho_p=\,n_p\,m_p=\rho_p^0 a^{-3(1-\nu)}$, where $\rho_p^0=n_p^0\,m_p^0$. Ditto, of course, for the (bound) neutrons. Therefore we find a possible scenario for the anomalous matter conservation law introduced in Eq.\,(\ref{mRGa1}).

The small coefficient $|\nu|\ll 1$ tracing the anomaly in the mass conservation law above can be identified with the parameter we have used in the previous sections, which can be attributed the meaning of $\beta$-function coefficient for the running of the CC. In fact, the same parameter must be involved in the induced time evolution of the vacuum energy density $\rL$, see e.g. Eq.\,(\ref{CRGa1}), and this is what makes consistent to have time evolving masses in General Relativity, i.e. provided the fundamental parameters of the gravitational sector change correspondingly such that the Bianchi identity remains preserved. Needless to say, in this framework not only the masses of the nucleons are slowly changing with the cosmic expansion, also the masses of all the nuclei in the universe\,\cite{FritzschSola2012}!

Furthermore, being the matter content of the universe
dominated by the dark matter, we cannot exclude that these particles
also vary with cosmic time in a similar way, although perhaps with a different anomaly index $|\nu_X|\ll 1$, such that
\begin{equation}\label{eq:nonconservedmX}
m_X(a)=m_X^0\,a^{3\nu_X}\, \,, \ \ \ \ \ (|\nu_X|\ll 1)\,.
\end{equation}
In general the time evolution of the masses should be connected with both the dynamical vacuum energy and the evolution of the gravitational coupling $G$, precisely as indicated in Eq.\,(\ref{dGneqo}).  For example, for an isotropic and homogeneous matter fluid with density $\rho_m$ in the matter dominated epoch, it implies the following generalized conservation law:
\begin{equation}\label{BianchiGeneral2}
\frac{G'(a)}{G(a)}\,\left[\rmr(a)+\rL(a)\right]+{\rho}'_m(a)+\rho'_{\CC}(a)+\frac{3}{a}\rmr(a)=0\,,
\end{equation}
where primes again denote $f'(a)=df/da$. We recognize Eq.\,(\ref{dGneqo2}) as a particular case of (\ref{BianchiGeneral2}) for the case $\rL=$const.

Clearly, if masses change with the cosmic scale time as in (\ref{eq:nonconservedmp}) and/or (\ref{eq:nonconservedmX})  the above conservation law (\ref{BianchiGeneral2}) entails in general a simultaneous cosmic evolution of  $\rL$ and $G$. Thus, if experiments would finally confirm the time evolution of masses\,\cite{FundConst}, this could be a possible scenario to explain the origin of the dark energy in the universe!

\section{Vacuum energy and the cosmological constant problem}\label{sect:vacuum and CCP}

Since the proclaimed discovery of the Higgs boson at the CERN LHC collider\,\cite{HiggsDiscovery}, the issue of the vacuum energy can no longer remain hidden under the rug. Some form of vacuum energy is there, in fact a huge value in the electroweak theory! We have to give an answer to this question. The Brout-Englert-Guralnik-Hagen-Higgs-Kibble mechanism for spontaneous symmetry breaking (SSB)\,\cite{HiggsMechanism64}  is a highly successful one from the point of view of particle physics, but at the same time is at the root of the extremely pressing vacuum problem in the SM. Can we make it compatible with cosmology? Or are we keep giving elusive answers once again? Admittedly it is not easy to give a sensible answer to a huge conundrum like this, but some ideas have been put forward.

\subsection{Adjusting the vacuum energy to a tiny value}

The use of the vacuum energy to explain the dark energy in our universe is usually criticized on account of the important inherent difficulties to implement this concept from quantum field theory into the cosmological framework. While the difficulties are enormous, some of the most usual criticisms are not completely justified. This is so not only because an alternative solution has never been produced from another more solid perspective, but also because the vacuum contribution itself probably needs to be interpreted in a way different from the traditional one leading to the huge fine tuning. We will comment briefly in the next subsection about that possible re-interpretation.

Concerning the alternative candidates, there is a long list of dark energy proposals in the market\,\cite{CCproblem}. For simplicity we shall only mention here, just in passing, the use of fundamental quintessence\,\cite{Quintessence} fields. Honestly speaking, it should not be regarded as a better solution to the cosmological constant problem than the $\CC$-term itself. It is well-known that quintessence models are plagued with fine tuning problems of no less severity than the vacuum energy in its traditional formulation.  The very notion of CC in such context becomes degraded: the CC term $\CC$ (and with it the entire contribution to the vacuum energy from the SM fields) is declared to be nonexistent by fiat! It is suddenly replaced by some scalar field $\varphi$ which generates a non-vanishing DE density $\rho_{\varphi}$ from the sum of its potential and kinetic energies at the present time:
\begin{equation}\label{Vphi}
\rLo\to \rho_{\varphi}^0=\{(1/2)\kappa\,\dot\varphi^2+V(\varphi)\}_{t=t_0}\,.
\end{equation}
The coefficient $\kappa$ is in principle arbitrary, but for $\kappa<0$ the field is of phantom type. If the kinetic energy for $\varphi$ is small enough,  $\rho_{\varphi}^0\simeq \{V(\varphi)\}_{t=t_0}$ looks like an effective cosmological constant. The scalar field $\varphi$ is in
principle unrelated to the Higgs boson or any other field of the
SM of particle physics, or known extensions thereof; it is an entirely \textit{ad hoc} construct just introduced to mimic the cosmological term. The quintessence approach, though, is simple enough for triggering quick calculations and has largely influenced the usual parameterizations of the DE.

Quintessence, however, is just an amenable spin-off of the older (and much more ambitious) idea
aiming at a scalar field
capable of dynamically adjusting the vacuum energy to zero or, in fact, to the ``tiny'' number $\rLo\sim 10^{-47}$ GeV$^4$\,\cite{Dolgov82,OldDynamAdjust,PSW}\footnote{Needless to say, ``tiny'' does not mean anything for a dimensionful quantity. When one speaks of the ``very small' value of the cosmological constant or vacuum energy density, $\rL$, we are implicitly comparing it to the fourth power of the mass of the average particle in the SM of strong and electroweak interactions, say $\sim 1$ GeV$^4$ at least. In actual fact only a very light ($m_{\nu}\sim 10^{-3}$ eV) neutrino would do. But not even this particle would be a strict SM candidate! And what about the remaining particles? Take the top quark, for instance, or the mass of the purported Higgs boson found at the LHC, and check the corresponding number $\sim m^4$. Hopes for a sensible answer evanesce fast!}. It is only in more recent times that this approach took the current popular form and was mainly applied to explain the possible dynamical character of the DE, with an eye at solving the ``cosmic coincidence'' problem. Quintessence, therefore, focuses exclusively on the question of why the current cosmological parameters
$\OL$ and $\OM$ are both of order $1$ despite the matter density is continuously decreasing with time. Notwithstanding, quintessence has nothing to say about the value of the cosmological term, i.e. on the ``old CC problem''\,\cite{CCproblem}.

Actually, quintessence is not free itself from serious problems (apart from its completely \textit{ad hoc} nature!). The field $\varphi$ is usually thought of as being part of some GUT which is operative at a high scale $M_X\sim 10^{16}$ GeV, or even at the Planck scale $M_P\sim 10^{19}$ GeV. It is very difficult to understand how the dynamics of the field $\varphi$ can possibly know about the comparatively tiny mass scale associated to the measured value of the
cosmological constant, namely the scale $m_{\CC}\equiv\left(\rLo\right)^{1/4}\sim 10^{-3}$ eV. As a result quintessence is doomed into a severe fine tuning problem, no less disturbing than the traditional approach to the vacuum energy!

Furthermore, the typical mass of the quintessence field should be of the order of the Hubble parameter now: $m_{\varphi}\sim\,H_0\sim 10^{-33}\,eV$, meaning a particle mass $30$ orders of magnitude below the very small mass scale $m_{\CC}\sim 10^{-3}$ eV we are trying to explain in the CC problem. This is kind of preposterous, isn't it? One may wonder if by admitting the existence of an ultralight
field like $\varphi$ (totally unrelated to the rest of the particle
physics world) is not just creating a problem far more worrisome
than the CC problem itself!

Of course many more ideas have been put forward to try to ameliorate the situation with the vacuum energy problem\,\cite{CCproblem}, see e.g. some recent attempts of very different nature \cite{Padmanabhans,Bennie2013,Barrow2011ab}, and also the review\,\cite{JSP-CCReview13} and references therein. Let us also mention the proposed especial class of modified gravity models considered in \cite{RelaxedUniverse,RelaxedFit,RelaxedBauer}, which proved efficient to technically implement a dynamical mechanism for relaxing the cosmological vacuum energy to a very small value starting from an arbitrary one in the early universe. This goes well beyond (!) the usual modified gravity approach where the DE is treated as a late-time effect\,\,\cite{ModifiedGrav}. The relaxation mechanism can actually be extended to the astrophysical domain so as to e.g. reduce the vacuum energy in the Solar System to a completely harmless level, see\,\cite{RelaxedSolaSystem} for details.

\subsection{Renormalizing away the vacuum energy}

Despite the aforementioned works showing the technical possibility to counteract the highly undesirable effects of a value of the vacuum energy comparable to the typical one predicted in ordinary QFT, one can adopt a different point of view that could prehaps shed some light to this conundrum\,\cite{JSP-CCReview13}. Rather than trying to adjust the value of the total vacuum energy density in the universe, we may take the point of view that the vacuum energy actually does {\emph not} produce any effect, if there is no change in the dynamical conditions. Instead of absorbing from the start the $\CC$-term on the \textit{r.h.s.} of Einstein's equations (\ref{EEsplit}),  we may imagine that the geometry effect of $\CC$ on the \textit{l.h.s.}  is exactly compensated by the corresponding energy effect $8\pi G\rL$ on its \textit{r.h.s.} Formally, this can be justified in QFT as a renormalization effect. Take for the moment flat Minkowskian spacetime. The {\em physical} vacuum energy density must be zero in it, otherwise we could not satisfy Einstein's equations in that geometry. The two contributions that add up to zero can be viewed as splitting ``zero'' into a UV-divergent effect, e.g. the result of the zero-point energy contribution of the fields (which goes on the  \textit{r.h.s.} of Einstein's equations) plus the bare contribution (which stays from the beginning on the \textit{l.h.s.}). The bare contribution is the original parameter in the Einstein-Hilbert geometric action, and in a QFT context with quantized matter fields, it can be split into a renormalized part (in some appropriate renormalization framework) plus a counterterm (adapted to that renormalization scheme):
\begin{equation}\label{counterterm}
\rL^b=\rL(\mu)+\delta\rL.
\end{equation}
The counterterm $\delta\rL$ must be chosen such that UV divergences cancel, and not even finite parts remain, such that the physical vacuum energy density is just zero in Minkowskian spacetime. Some subtleties need to be addressed here\,\cite{JSP-CCReview13}, of course, but let us now move to the curved spacetime case.

Recall that the Cosmological Principle is based on the homogeneous and isotropic FLRW line element
\begin{equation}\label{eq:FLRWmetric}
ds^2=c^2dt^2-a^2(t)\left[\frac{dr^2}{1-K\,r^2}+r^2\,d\Omega^2\right]\,,
\end{equation}
where $K$ is the spatial curvature parameter.  As we know, $K\simeq 0$ is strongly favored  by the observations, and $K=0$  is actually required (to within exponential precision) by the inflationary paradigm as a means to solve important cosmological problems -- such as the horizon problem and the curvature problems.  We want nevertheless to keep $K$ for a while at this point, just for the sake of the following argument.

Solving Einstein's equations in this metric produces the two basic equations.  On the one hand we have Friedmann-Lema\^{\i}tre's equation:
\begin{equation}\label{eq:FriedmannK}
H^2\equiv\left(\frac{\dot{a}}{a}\right)^2=\frac{8\pi
G}{3}\rmr+\frac{\CC}{3}-\frac{K}{a^2}\,,
\end{equation}
where the constant $K$ is the spatial curvature parameter appearing in
(\ref{eq:FLRWmetric}), and on the other the acceleration equation:
\begin{equation}\label{eq:acceleration}
\frac{\ddot{a}}{a}=-\frac{4\pi\,G}{3}\,(\rmr+3\pmr)+\frac{\CC}{3}\,.
\end{equation}
Equations (\ref{friedr}) and (\ref{friedr2}) are a particular case of the previous equations for $K=0$, as readily seen using $\ddot{a}/a=H^2+\dot{H}$.

Let us recall that the total spacetime curvature of the universe within the FLRW metric is given by
\begin{equation}\label{Rvalue}
R=-6\,\left(\frac{\ddot{a}}{a}+\frac{\dot{a}^2}{a^2}+\frac{K}{a^2}\right)\,,
\end{equation}
where $K$ is the spatial curvature.

Substituting (\ref{eq:FriedmannK}) and (\ref{eq:acceleration}) into Eq.\,(\ref{Rvalue}) we find:
\begin{equation}\label{Rscalar}
R=-8\pi\,G\,\left(4\rL+\rmr-3\,p_m\right)=-4\,\CC-8\pi\,G\left(1-3\wm\right)\rmr\,,
\end{equation}
where $\wm=p_m/\rmr$ is the EoS of matter.
We can easily check that this equation is also obtained directly from Einstein's equations (\ref{EEsplit}) or \,(\ref{EE}) and the explicit form of the energy momentum tensor (\ref{Tmunuideal}). Contracting the indices we immediately get $R=-8\pi\,G\tilde{T}=-8\pi\,G\left(4\rL+T\right)$ where $T\equiv T^{\mu}_{\mu}=\rmr-3p_m$ is the trace of the energy momentum tensor for matter. Hence we arrive once more at Eq.\,(\ref{Rscalar}).

From Eq.\,(\ref{Rvalue}) we see that a static universe ($\dot{a}=\ddot{a}=0$) without spatial curvature ($K=0$) has vanishing spacetime curvature: $R=0$. Hence it must necessarily be Minkowskian spacetime. On the other hand from Eq.\,(\ref{Rscalar}) we see that such a spacetime must be fully empty, that is to say, it must have zero matter content. Not only so, it must also have exactly zero vacuum energy. Hence, $\rmr=\rL=0$!

From these simple observations it is clear that QFT in Minkowski spacetime is, strictly speaking, inconsistent with General Relativity unless we assume that there is no matter and vacuum energy whatsoever. So even if we assume that we are in the vacuum state of QFT we must have $\rL=0$.  In contrast, in usual calculations in Minkowski spacetime we naively find that this is not the case, e.g. when we compute the zero-point energy (ZPE) carried by any of the SM fields. We can (in fact, {\em we must}) also add here the vacuum energy density induced by the Higgs mechanism of the SM, if we believe in it. For example, the tree-level contribution can be written in terms of the two measurable quantities, $M_H\simeq 126$ GeV (physical Higgs mass\,\cite{HiggsDiscovery}) and $G_F\simeq 1.166\times 10^{-5}$ GeV$^{-2}$ (Fermi's constant), as follows\,\cite{JSP-CCReview13}:
\begin{equation}
\langle V \rangle=-\frac{1}{8\sqrt{2}}\,\frac{M_{H}^2}{G_F}\simeq -1.2\times 10^{8}\,{\rm GeV}^4\,.
\label{eq:Higgstree}
\end{equation}
This value is some $55$ orders of magnitude out of range (and opposite in sign) as compared to to the tiny measured value $\rLo\sim 10^{-47}$ GeV$^4$ of the cosmological constant (expressed as a vacuum energy density).

If that is not enough, all these calculations should be pushed to very high orders in perturbation theory owing to the extremely small value $\rLo$ as compared to the average mass scale in the SM. An estimate \cite{JSP-CCReview13} says that loop diagrams of order 20th may still be contributing!

The situation with the ZPE contribution is not better\,\cite{JSP-CCReview13}. A simple calculation at one-loop order for a real scalar field shows that the effect is, after renormalization, proportional to the fourth power of the particle mass, i.e. $\sim m^4$. For example, in the $\overline{\rm MS}$ scheme in dimensional regularization one finds\,\cite{ShapSol00,Akhmedov2002}:
\begin{equation}\label{Vacfree2}
V_{\rm ZPE}^{(1)}=
\frac12\,\mu^{4-n}\, \int\frac{d^{n-1} k}{(2\pi)^{n-1}}
   \,\sqrt{{\bf k}^2+m^2}
= \frac12\,\beta_\Lambda^{(1)}\,\left(-\frac{2}{4-n}
-\ln\frac{4\pi\mu^2}{m^2}+\gamma_E-\frac32\right) \,,
\end{equation}
where $n$ is the spacetime dimension, $\gamma_E$ is Euler's constant, and
\begin{equation}
\label{beta4} \beta_\Lambda^{(1)}=\frac{m^4}{2\,(4\pi)^2}
\end{equation}
is the one-loop $\beta$-function for $\rL$ in that scheme\,\cite{ShapSol00}. It contains the term that carries the nasty factor $\sim m^4$.

For the average particle in the SM a contribution to $\rL$ of order $\sim m^4$ is completely out of range (see the last footnote). Of course the above result (\ref{Vacfree2}) is still an unrenormalized one since it diverges for $n\to 4$. A detailed renormalization in curved spacetime is complicated and will not be discussed here\,\cite{BooksQFT} -- see \cite{JSP-CCReview13} for a short account --, but we can single out what is the fate of the $m^4$ terms in the (conventionally) renormalized result. As a matter of fact they stay in the final outcome for curved background, and cannot be removed if we follow the normal choice of the counterterm $\delta\rL$ mentioned in Eq.\,(\ref{counterterm}). Indeed, if we choose the latter just to cancel the pole at $n=4$, and maybe some additive term, we find:
\begin{equation}\label{renormZPEoneloopCurved}
\rho_{\rm ZPE}^{(1)}=\rL(\mu)+\frac{m^4}{4\,(4\,\pi)^2}\,\left(\ln\frac{m^2}{\mu^2}+{\rm finite\ const.}\ \right)\,.
\end{equation}
The additive constant term is unimportant here. What matters is the obvious presence of the $\sim m^4$ factor in the renormalized result. Needless to say, the above result is independent of $\mu$ since $\rL(\mu)$ (the renormalized vacuum parameter) is also $\mu$-dependent whereas the final ZPE result is overall independent of the renormalization scale.

It is ``obvious'' that we cannot cure the CC problem by some miraculous cancelation between the various sources of vacuum energy in the SM, unless we use dynamical mechanisms of the sort we have mentioned in the previous section. But for the time being none of these mechanisms has in fact a fundamental theory behind.
Another possibility, as mentioned in the beginning of this section, is that we can simply renormalize away the vacuum energy in QFT. Rather than hoping for miraculous cancelations, we can just proclaim a renormalization condition\,\cite{JSP-CCReview13} much in the same way as we renormalize, say, the electron mass in the on-shell scheme.

Here we can apply a renormalization setting precisely at a point in metric space where we know for certain, as discussed previously, that the CC is exactly zero: Minkoswski spacetime. If we restrict to the ZPE case at one-loop, this leads to fix the counterterm as follows:
\begin{equation}\label{deltaMSBPhysical}
\rL(\mu)+\delta\rL+\frac{m^4\,\hbar}{4\,(4\pi)^2}\,\,\left(-\frac{2}{4-n}
-\ln\frac{4\pi\mu^2}{m^2}+\gamma_E-\frac32\right)=0\,.
\end{equation}
With this setting we no longer obtain the troublesome Eq.\,(\ref{renormZPEoneloopCurved}), we just obtain ``zero value'' for the physical ZPE (at one-loop). Of course the same setting should be extended at arbitrary order in perturbation theory and we can include the contribution of the Higgs potential at the given order of perturbation theory. Notice that we would do exactly the same with the electron mass: we would split its bare value as $m_b=m_e+\delta m$ and we would choose $\delta m$ so as to cancel the radiative corrections at the given order of perturbation theory  such that the measured value of the electron mass, $m_e$, stays unscathed. In practice things are more complicated in the cosmological case\,\cite{ShapSol09}. We should start from a more physical renormalization framework such that the calculation becomes closer to the simple on-shell treatment of the electron mass case. This is, as it may be suspected, more easily said than done!

However, the basic QFT principle is the same in all cases: by imposing that the infinite tower of radiative corrections will not change an inch the experimentally measured value of a physical quantity, we cannot speak of fine tuning, we call it renormalization of the corresponding physical quantity, whether it is the electron mass or the vacuum energy density in Minkowskian spacetime.  Many formal techniques are possible also in QFT in curved spacetime, see e.g.\,\cite{Tommi} for a recent approach. Fine tuning, in contrast, is a horrendous numerical imbalance that appears only when we play with an assortment of finite quantities at different orders of magnitude. Then we stumble upon unsurmountable difficulties, such as to try to make sense of the value of the cosmological constant after considering the many phase transitions through which the universe evolved from the very early times till the present days.

Treating instead the problem as renormalization makes it not different from the ordinary situation in QFT. We can, then, forget about the role played by the finite terms. After all there is always an arbitrary finite term involved in the renormalization procedure, which is fixed by comparison with observation. Take the ZPE source of vacuum energy, which we know is purely a quantum effect (originating from the vacuum-to-vacuum diagrams to all orders) plus the Higgs contribution which has a classical plus a quantum part. We subsequently impose a renormalization condition that involves as usual UV-divergent terms along with finite parts, and the physical quantity that remains is the neat zero value of the vacuum energy in Minkowskian spacetime, valid at all orders of perturbation theory.

The following observation is now in order. One could object that in the electron case the dependence is logarithmic in $m_e$ (and cutoff) whereas the renormalization condition (\ref{deltaMSBPhysical}) strongly clashes with the hierarchy problem in a way similar to the unnatural situation of the Higgs boson mass in the GUT context, being even worse if we take into account the quartic dependence of the mass term in the CC case. This observation would be justified if $\rL$ were a rigid quantity, but as we know it is not so in this context since $\rL$ takes on the quartic power of $H$ in the early universe, see Eq.\,(\ref{lambdaH2H4}). It follows that if we establish the renormalization condition (\ref{deltaMSBPhysical}) at a scale near $H\sim m$ (where $\rL$ was of order $H^4\sim m^4$) it becomes perfectly natural. Put another way, such renormalization condition is naturally formulated only in the early universe, not in the current one where the typical energy scale defined by the Hubble parameter is far below the average mass of the SM particles\,\cite{JSP-CCReview13}. Doing otherwise leads us to the puzzling hierarchy problem. Such problem could not be avoided in the presence of a traditional $\CC$-term, which is precisely the situation with the $\CC$CDM model. Therefore, only for a dynamical vacuum term a natural renormalization condition can be formulated!

Once that renormalization condition has been imposed for Minkowski space -- i.e. the geometry where we know it must hold exact --  the $\sim m^4$ terms will no longer appear, they are banned forever. What is left in the expanding spacetime case is a quantum effect of curvature. This effect is distinctive with respect to the Minkowski case much in the same way as is distinctive a change of the boundary conditions in a Casimir device. In the Casimir effect\,\cite{Casimir} what one measures indeed is the vacuum configuration differences of the system with respect to free Minkowski space. For discussions in different frameworks, see e.g. \cite{Casimir2,Casimir3,Zhitnitsky,Elizalde2014}.

Similarly, in the cosmological situation we are comparing the expanding universe with Minkowski spacetime and we can read off the differential response of the system. It is characterized by the curvature effects carried by the terms on the right hand side of equations (\ref{seriesRLH}), (\ref{lambdaH2H4}) or (\ref{QFTModelLowEnergy}) --- depending on the epoch of the universe under study. This is what justified the analysis of the cosmic evolution of the vacuum energy in the previous sections. We did not care there about the seemingly huge contributions from the ZPE of the quantum fields, or from the Higgs potential and its multifarious quantum corrections, simply because we renormalized them away once and for all for Minkowski space\,\cite{JSP-CCReview13}. What is left in the expanding universe, as a net effect, is the contribution from the aforementioned terms, which truly make the vacuum energy be a function of $H$ and possibly also of $\dot{H}$, i.e. $\rL(H,\dot{H})$. Once more we only measure differences, not totals. The bulk of these totals cancels out on each side of Einstein's equations after appropriate renormalization, as explained in the beginning.

\section{Summary}

We have discussed models with time evolving vacuum energy $\rL(t)$ (including in some cases also variable gravitational coupling G) and assessed their impact in the cosmic history. We have emphasized that dynamical vacuum energy in an expanding universe is expected within QFT in curved spacetime\,\cite{JSP-CCReview13}. We have paid particular attention to models of running vacuum energy potentially connected with the possible quantum effects on the effective action of QFT in curved spacetime.  Specifically, we have shown that if the vacuum energy density evolves as a function of the form
\begin{equation}\label{eq: GeneralRG}
\rL(t)=c_0+\sum_{k=1} \alpha_{k} H^{2k}(t)+\sum_{k=1} \beta_{k}\dot{H}^{k}(t)\,,
\end{equation}
with $c_0\neq 0$ (viz. an ``affine'' function constructed out of  powers of $H^2=\left(\dot{a}/a\right)^2$ and $\dot{H}=\ddot{a}/a-H^2$, hence with an even number of time derivatives of the scale factor $a$), one can formulate a unified model of the cosmological evolution, compatible with the general covariance of the effective action, in which inflation is predicted, a correct transition (``graceful exit'') into a radiation phase can be naturally accommodated, and finally the late time cosmic evolution can also be successfully described.

\vspace{0.25cm}

These kind of models, if we reparameterize them as a self-conserved DE fluid at fixed G (typically at $\CC=0$, as done in quintessential philosophy), generally lead to a non-trivial effective equation of state.  In this picture, which is the common one in most studies of the dark energy, the running models can effectively appear as quintessence, and even as phantom energy or a mixture of both. This feature  can help explaining the observation of possible ``mirage transitions'' from quintessence to phantom, or vice versa, without invoking at all the existence of fundamental quintessence or phantom scalars fields. Our analysis serves also to illustrate the general fact that parametric degeneracies (or, as they are also called, cosmographic degeneracies) along with the unknown nature of dark energy makes it very tricky to work with parametric forms of the equation of state of dark energy fitting cosmological data. It turns out that the running vacuum models under consideration cannot be fitted to the most common forms of the equation of state of the dark energy. These are usually very simple, but prove unable to capture essential traits of more realistic models.

\vspace{0.25cm}
The recent observation of a Higgs-like boson particle at CERN's LHC\,\cite{HiggsDiscovery} reminded us that
the vacuum energy (through the spontaneous symmetry breaking mechanism)
can be a fully tangible concept in real phenomenology. Despite the usual
difficulties with this concept, the Higgs discovery points once more to the notion of vacuum energy in quantum field theory. It should spur us to understand its meaning more deeply, rather than keep on blindly replacing it by a variety of other concepts, each one more exotic and fathomless than the previous.

In the meanwhile the effects associated to changes in the vacuum energy can be as real as they are in, say, the Casimir effect, which is sensitive only to tiny changes of the vibrational modes of the vacuum configuration. The possible analogy suggests that in the cosmological case the changes in the vacuum energy density are described in terms of mildly evolving functions of the expansion rate, specifically in the form of powers of $H^2$ and $\dot{H}$, both being of order of the observed $\CC$. These functions describe the time evolution of the spacetime curvature of our expanding universe and can ultimately be responsible for the currently observed cosmic acceleration. As in the Casimir effect, these terms are tracing the differences of the vibrational modes of the curved geometry of the expanding vacuum as compared to the Minkowski one, after we have renormalized away the bulk vacuum energy of the latter\,\cite{JSP-CCReview13}.  What is left is the measurable quantity that we may unnecessarily have called ``dark energy'' in the physical spacetime.

\vspace{0.25cm}

As pointed out in our discussion, the ultimate origin of the ``dark energy'' might be connected to the covariant response of General Relativity impinged on the parameters of the gravitational sector, $\CC$ and $G$, to the possible time evolution of the particle masses in the universe as triggered by the dynamics of particle physics \,\cite{FritzschSola2012,FritzschSola2014}. If so, this could be revealing an intimate connection between the quantum physics of the micro and the macro Cosmos.

\begin{theacknowledgments}
I have been supported in part by FPA2010-20807 (MICINN), Consolider grant CSD2007-00042 (CPAN) and by DIUE/CUR Generalitat de Catalunya under project 2009SGR502. I am grateful to the organizers of the workshop \textit{Particle and Nuclear Physics at all scales, Astroparticle Physics and Cosmology}, held in Saint Petersburg, Russia, for the invitation and smooth organization of the conference, as well as for the partial support. It is a pleasure to thank S. Basilakos, H. Fritzsch, J.A.S. Lima and D. Polarski for the recent collaboration in some of the work presented here. Finally, I wish to express my gratitude to the Institute for Advanced Study at
the Nanyang Technological University in Singapore for the hospitality while the writing of these proceedings was being finished.
\end{theacknowledgments}


\newcommand{\JHEP}[3]{ {JHEP} {#1} (#2)  {#3}}
\newcommand{\NPB}[3]{{ Nucl. Phys. } {\bf B#1} (#2)  {#3}}
\newcommand{\NPPS}[3]{{ Nucl. Phys. Proc. Supp. } {\bf #1} (#2)  {#3}}
\newcommand{\PRD}[3]{{ Phys. Rev. } {\bf D#1} (#2)   {#3}}
\newcommand{\PLB}[3]{{ Phys. Lett. } {\bf B#1} (#2)  {#3}}
\newcommand{\EPJ}[3]{{ Eur. Phys. J } {\bf C#1} (#2)  {#3}}
\newcommand{\PR}[3]{{ Phys. Rep. } {\bf #1} (#2)  {#3}}
\newcommand{\RMP}[3]{{ Rev. Mod. Phys. } {\bf #1} (#2)  {#3}}
\newcommand{\IJMP}[3]{{ Int. J. of Mod. Phys. } {\bf #1} (#2)  {#3}}
\newcommand{\PRL}[3]{{ Phys. Rev. Lett. } {\bf #1} (#2) {#3}}
\newcommand{\ZFP}[3]{{ Zeitsch. f. Physik } {\bf C#1} (#2)  {#3}}
\newcommand{\MPLA}[3]{{ Mod. Phys. Lett. } {\bf A#1} (#2) {#3}}
\newcommand{\CQG}[3]{{ Class. Quant. Grav. } {\bf #1} (#2) {#3}}
\newcommand{\JCAP}[3]{{ JCAP} {\bf#1} (#2)  {#3}}
\newcommand{\APJ}[3]{{ Astrophys. J. } {\bf #1} (#2)  {#3}}
\newcommand{\AMJ}[3]{{ Astronom. J. } {\bf #1} (#2)  {#3}}
\newcommand{\APP}[3]{{ Astropart. Phys. } {\bf #1} (#2)  {#3}}
\newcommand{\AAP}[3]{{ Astron. Astrophys. } {\bf #1} (#2)  {#3}}
\newcommand{\MNRAS}[3]{{ Mon. Not. Roy. Astron. Soc.} {\bf #1} (#2)  {#3}}
\newcommand{\JPA}[3]{{ J. Phys. A: Math. Theor.} {\bf #1} (#2)  {#3}}
\newcommand{\ProgS}[3]{{ Prog. Theor. Phys. Supp.} {\bf #1} (#2)  {#3}}
\newcommand{\APJS}[3]{{ Astrophys. J. Supl.} {\bf #1} (#2)  {#3}}

\newcommand{\Prog}[3]{{ Prog. Theor. Phys.} {\bf #1}  (#2) {#3}}
\newcommand{\IJMPA}[3]{{ Int. J. of Mod. Phys. A} {\bf #1}  {(#2)} {#3}}
\newcommand{\IJMPD}[3]{{ Int. J. of Mod. Phys. D} {\bf #1}  {(#2)} {#3}}
\newcommand{\GRG}[3]{{ Gen. Rel. Grav.} {\bf #1}  {(#2)} {#3}}

\bibliographystyle{aipproc}   

\end{document}